\documentclass[a4paper,10pt]{article} 

\usepackage{enumitem}
\usepackage[linesnumbered,ruled,vlined]{algorithm2e}
\usepackage{algpseudocode}
\usepackage{etoolbox}
\usepackage{caption}
\usepackage{xcolor,colortbl} 
\usepackage{multirow}
\usepackage{booktabs}
\usepackage[titletoc,title]{appendix}
\usepackage[a4paper,
            left=1.2in,
            right=1.2in,
            top=1.4in,
            bottom=1.4in,
            ]{geometry}
\usepackage{graphicx}
\usepackage{amsmath}
\usepackage{amsfonts}
\usepackage[hidelinks]{hyperref}
\usepackage[onehalfspacing]{setspace}

\title{Investigating Similarities Across Decentralized Financial (DeFi) Services}

\author{Junliang Luo
\thanks{McGill University, Montréal, Canada. junliang.luo@mail.mcgill.ca}
\and Stefan Kitzler
\thanks{Complexity Science Hub Vienna, Vienna, Austria. kitzler@csh.ac.at} \,
\thanks{AIT Austrian Institute of Technology, Austria}
\and Pietro Saggese
\thanks{IMT School for Advanced Studies, Lucca, Italy. pietro.saggese@imtlucca.it}  $^\dagger$}

\makeatletter
\renewcommand{\@algocf@capt@plain}{}
\makeatother
\SetAlgorithmName{}{}{}

\begin{document}

\maketitle

\begin{abstract}

We explore the adoption of graph representation learning (GRL) algorithms to investigate similarities across services offered by Decentralized Finance (DeFi) protocols.
Following existing literature, we use Ethereum transaction data to identify the DeFi \textit{building blocks}. These are sets of protocol-specific smart contracts that are utilized in combination within single transactions and encapsulate the logic to conduct specific financial services such as swapping or lending cryptoassets. 
We propose a method to categorize these blocks into clusters based on their smart contract attributes and the graph structure of their smart contract calls. 
We employ GRL to create embedding vectors from building blocks and agglomerative 
models for clustering them. 
To evaluate whether they are effectively grouped in clusters of similar functionalities, we associate them with eight financial functionality categories and use this information as the target label.
We find that in the best-case scenario purity reaches $.888$.
We use additional information to associate the building blocks with protocol-specific target labels, obtaining comparable purity ($.864$) but higher V-Measure ($.571$); we discuss plausible explanations for this difference. 
In summary, this method helps categorize existing financial products offered by DeFi protocols, and can effectively automatize the detection of similar DeFi services, especially within protocols. 

\end{abstract}

\section{Introduction}

Decentralized Finance (DeFi) refers to a novel financial paradigm that leverages self-executing code deployed on top of Distributed Ledger Technologies (DLTs), known as smart contracts, to provide financial functionalities within a decentralized framework. Thereby, it eliminates the need for intermediary entities like centralized financial institutions for transaction settlement.

The DeFi ecosystem is a thriving environment for financial innovation and the conception of new financial products~\cite{auer2023technology}. Automated market-making, for instance, is a mechanism that facilitates the decentralized trading of cryptoassets~\cite{xu2023sok} by substituting order books. The interoperability of smart contracts enables the creation of `DeFi compositions', where financial services of several DeFi protocols are combined to offer novel, complex and deeply nested financial products~\cite{kitzler2022systematic}. %
The DeFi landscape is indeed evolving rapidly: permissionless DLTs are censorship-resistant and their open-source design enables everyone to create new financial products and protocols; to date, it is hard to keep track of all existing protocols and the financial services they offer. 
Notwithstanding, such services often overlap in scope and purpose: most of them enable the lending and trading of cryptoassets, or devise yield-bearing strategies to reward liquidity provision. 
When protocol developers deploy newer versions of their projects, for interoperability purposes it is often essential that the newly deployed smart contracts are compatible with previous versions.
It is also well known that code reuse is a practice conducted by protocol developers; as an example, the protocol Sushiswap is a fork of Uniswap~\cite{fan2023towards}. Therefore, it is not uncommon that the implemented functionalities are similar if not identical across protocols. 
More generally, DeFi services reproduce in a decentralized context the functional logic of established `traditional finance counterparts'. One could expect that the logic of such functions shares some fundamental characteristics.
In other words, it is likely that different implementations of similar financial services share similar attributes and structure. 

Recent research has disentangled the different financial services produced by DeFi protocols and has identified fundamental sets of protocol-specific smart contracts that, in combination, encapsulate the logic to conduct specific financial functionalities, such as swapping cryptoassets, or lending and borrowing them, calling them the \textit{building blocks} of DeFi~\cite{kitzler2022disentangling}. These building blocks are defined univocally by their shape and structure, which is modelled as a tree-like structure of interacting smart contracts.
However, previous research did not investigate in detail the similarities and differences across building blocks, and to the best of our knowledge, no other studies have investigated in depth the entire space of such DeFi services~\cite{khan2022graph}. 

For the aforementioned reasons, we believe it is important to conduct a more detailed analysis of the various financial functions present within the DeFi ecosystem.
On the one side, the growth without oversight of the DeFi ecosystem serves as a motivation for studies exploring solutions to automate the categorization of DeFi services; on the other side, evidence of similitudes on a technical level and of code reuse motivates studies that aim at investigating such similarities. 

In this paper, we exploit machine learning algorithms to investigate how well it is possible to categorize building blocks in clusters of similar financial functionalities, based on their graph structure and attributes. 
We also aim to explore whether the financial functionalities of building blocks can be identified by their location within an embedding space or their proximity to certain other building blocks.
Finally, we aim to understand the specific common design patterns that lead to the formation of these clusters and analyze imperfect classifications within them.

To answer these questions, we first replicate existing methods~\cite{kitzler2022disentangling} to obtain a set of building blocks encapsulating the main financial functions of DeFi. 
To assess the similarity between these building blocks, we produce a similarity embedding space by applying graph-level representation learning and then exploit agglomerative clustering models on building blocks. We obtain four different specifications by associating various features to the nodes that compose the building blocks. Next, to evaluate how financial functionalities are grouped within such space, and to analyze whether distinct clusters represent particular financial operations, we gather additional information indicating the financial functionality category or the protocol they are associated with, and use it as the target label to compute a number of metrics such as homogeneity, completeness, V-Measure, and purity.

When evaluating the results using the `financial functionality category' label, we find that the outcomes yielded the highest value among all specifications for purity with .888, but a relatively low V-Measure (.239).
When evaluating the results using the `protocol' label, we find that the values for purity are comparable (.864 for the best-performing specification), and higher V-Measure (.571).
To explain the difference in performance, we investigate more closely the common patterns within protocol-specific building blocks and look for plausible explanations. Finally, we identify protocol-specific patterns, that are re-used across them in different financial functionalities, which likely explain
higher values for the clustering evaluated on the protocol target labels, compared to the financial functionalities.

The paper is structured as follows. Section \ref{sec:bg} describes in detail the concepts of building blocks and DeFi compositions, and discusses the related literature.  In Section \ref{buildling_block_embedding} we describe the data used and the methodology, while in Section \ref{sec:clustering} we show the results. Finally,  in Section \ref{sec:conclusions} we discuss our findings and report our concluding remarks.

\section{Background}
\label{sec:bg}

\subsection{Decentralized Finance \& DeFi Protocols}

DeFi aims for open access for its users and provides a decentralized ecosystem that does not need intermediaries such as financial institutions to settle transactions. 
To date, Ethereum is the main blockchain for DeFi. In contrast to UTXO-based cryptocurrencies, such as Bitcoin, the innovation of the Ethereum Virtual Machine (EVM) enables the use of Contract Accounts (CAs), which are basically software programs deployed on the blockchain.
In contrast to Externally Owner Accounts (EOAs), CAs contain program code and, once deployed on the blockchain, methods (functions) can be called and the implemented logic will then be executed and computed.
The most popular implementation of CA is cryptoassets (later also just assets), representing real-world assets or rights on the blockchain.
Prominent examples are stablecoins, such as Tether or USDC, whose value is pegged to the US dollar.
DeFi can be thought of as an entire ecosystem of financial services for cryptoassets. The so-called DeFi protocols are implemented through CAs on the blockchain and provide financial functionalities, such as decentralized exchanges (DEX) or lending, to the end users. 
Many services automate their continuous code launch, creating factory-deployed (FD) contracts.
Previous literature has extensively studied DeFi protocols, especially DEXs~\cite{Xu2022a, Parlour2021, Heimbach2022risks} using Automated Market Makers (AMM)~\cite{Fritsch2022}, Lending~\cite{Heimbach2023short, Sun2022a, Xu2022} or Derivatives~\cite{Xiong2023a} protocols.

\subsection{DeFi Compositions}\label{sec:DeFiComp}

Ethereum flexibility allows CAs, deployed on the blockchain, to be called by other CAs, therefore enabling smart contract interoperability. Consequently, entire DeFi protocols can leverage financial functionalities offered by other protocols. %
Such a combination of multiple DeFi protocols, also known as DeFi compositions, might be beneficial for automating or providing more sophisticated actions.
For instance, an aggregator protocol $P_i$ can be used to determine the DEX protocol $P_j$ with the best available price and execute the swap from Asset $A$ to Asset $B$ within the same, single transaction. Intuitively, this is a DeFi composition between $P_i$ and $P_j$. 
Technically, a DeFi composition can be thought of as
a combination of CA functionalities such that CAs associated to different DeFi protocols are called from another in one single transaction. A more thorough definition is given by Kitzler et al.~\cite{kitzler2022systematic}.

\begin{figure}[t]
\centering
\includegraphics[width=\columnwidth]{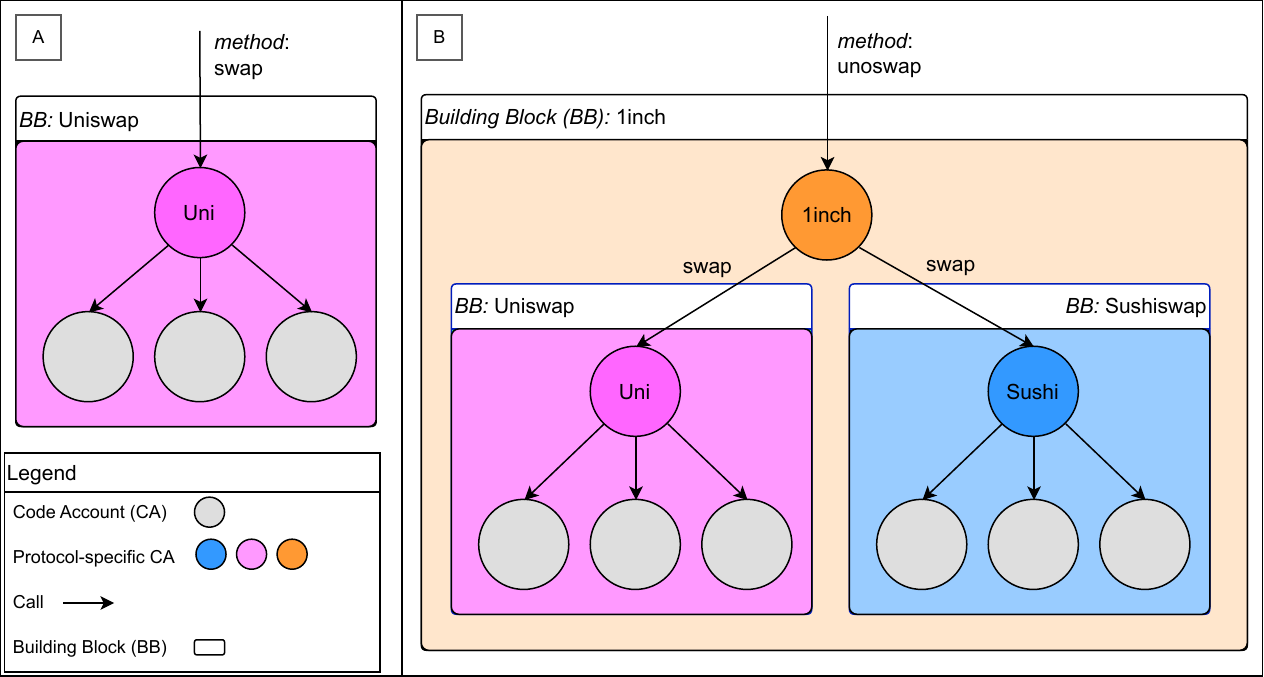}
\caption{Building blocks examples from \textit{Uniswap} and \textit{1inch}. Nodes represent account addresses and links are the calls of other accounts methods. Tree-structured transactions reveal the nested structure of composed DeFi services. }
\label{fig:bblks_examples}
\vspace{0.3cm}
\end{figure}

\subsection{Conceptualization of Building Blocks}\label{sec:bb}

Whilst interoperability increases the potentialities of DeFi, it also adds a level of complexity to the system. Previous research~\cite{kitzler2022disentangling} investigated DeFi transactions in Ethereum in order to identify DeFi compositions and disentangle them. The authors also proposed an algorithm that allows to identify sets of smart contracts that, in combination, encode the logic of fundamental financial functions, that they call building blocks.

Building blocks are functional units of DeFi protocols that can be derived from the call structure of single transactions and are constructed as follows. 
Beginning with an EOA as the initiator of the transaction, the call of a CA can lead to subsequent internal transactions, i.e. calls of other CAs' methods, and ultimately culminate in a cascading, tree-like transaction structure.
Building blocks can be extracted as sub-trees, where the root node (i.e., the top node of the tree structure) is a protocol-specific CA with no incoming but at least one outgoing link.
Furthermore, a building block may also encompass additional sub-trees, essentially forming nested structures of blocks.
Figure~\ref{fig:bblks_examples} provides graphic support to understand this concept. It represents two frequently appearing building blocks from the popular DEX \emph{Uniswap} (A) and one from the aggregator protocol \emph{1inch} (B). 
Nodes represent account addresses
and links represent the calls of CAs' methods.
It illustrates the tree-like structure of building blocks and reveals
the atomic \emph{swap} block of \emph{Uniswap} (A) to exchange assets. The more complex \emph{1inch} building block (B) executes a swap that derives the best price by leveraging both building block (A) and an equivalent \emph{Sushiswap} building block, as intuitively described in Section \ref{sec:DeFiComp}, therefore creating a DeFi composition.

\subsection{Related Work} 

Compositions of DeFi protocols increase the ecosystem complexity, which makes systemic risk even harder to assess~\cite{Schaer2021a, harvey2021defi, kitzler2022systemic, Werner:2021to, Jensen:2020uv}. 
Early works were conducted for decomposing other elements of DeFi but not across protocols. Moin et al. \cite{moin2019classification} decomposed the design of stablecoins into various component design elements and discussed the strengths and drawbacks. Tolmach et al. \cite{tolmach2021formal} specified systems compositions in automated market makers (AMMs) decentralized exchanges.
Von Wachter et al.~\cite{Wachter2021} described the `composed' derivatives of assets, in other contexts also known as `wrapped' tokens, and showed that the complexity of wrapping operations has been peaking in the third quarter of 2020, a period often referred
to as `DeFi Summer'.
Existing online tools, such as DeFiLlama\footnote{https://defillama.com/categories}, provide categories of protocols, but do not necessarily distinguish among financial functionalities offered.

Other studies related to our work are various machine learning-based analyses built upon extracting features of blockchain transactions to reveal the insights of the transactions in the blockchain blocks for assessing DeFi risks.
Methods were proposed to process the input transaction graphs to construct desired sub-structure graphs or newly derived graphs for further analysis.
Weber et al.~\cite{weber2019anti} explored the potential of Graph Convolutional Networks (GCN) for anti-money laundering (AML) in Bitcoin, contributing a publicly available labelled dataset of Bitcoin transactions for financial forensics.
Li et al.~\cite{li2022ttagn} introduced a graph neural network-based model that incorporates temporal transaction data for the identification of phishing scams on the Ethereum network.
Pocher et al.~\cite{pocher2023detecting} demonstrated employing Graph Convolutional Networks (GCN) and Graph Attention Networks (GAT) for modelling blockchain transactions as complex networks enhances the detection of AML and Countering the Financing of Terrorism (CFT) anomalies.
Han et al.~\cite{han2023mt} developed a multi-layer graph neural network-based model for temporal transaction anomaly detection within Ethereum's multi-token transaction networks.

\section{Building Block Embedding}
\label{buildling_block_embedding}
This section describes the dataset and the methodology devised to produce embeddings for DeFi building blocks. 
To uncover functional similarities and recurring interaction patterns, we first produce an embedding space where each building block is mapped to a high-dimensional vector by a graph representation algorithm; in this procedure, we assign various features to each node (i.e., smart contract) of each building block. 
Similarity searches are conducted based on their distance within the space.
We utilize two different target labels for evaluating the performance of our approaches in grouping DeFi building blocks into clusters based on their similarity.

\subsection{Dataset and Sources Description}
The dataset consists of the top 10,000 building blocks ranked by count and identified using the building block extraction algorithm, as proposed in Kitzler et al.~\cite{kitzler2022disentangling}. 
The building blocks are extracted utilizing all Ethereum transactions from January 1, 2021 (block 11,565,019) to August 5, 2021 (block 12,964,999) and involve 23 DeFi protocols and their contracts already identified in previous literature\footnote{Protocols: \textit{Badger, Convex, Fei, Harvestfinance, RenVM, Vesper, Yearn, Barnbridge, dYdX, Futureswap, Hegic, Nexus, Synthetix, 0x, 1inch, Balancer, Curvefinance, SushiSwap, UniSwap, Aave, Compound, Instadapp, Maker}. We note that the choice of the protocols, their associated contracts, and the timeframe are conditional to the availability of sources published by the authors of previous studies.}.
The algorithm extracts the building blocks by mapping the transactions into edge-induced subtrees, subsequently hashed based on a combination of their execution order, target nodes' outdegrees, and associated method.
Each building block in the dataset has fields including the executed method's name, subtraces that outline the tree structure and participating addresses of the transactions within a building block, and a count that reflects the frequency of the building block's occurrence.

\subsection{Methods of Producing Embedding}
Given the building block tree-shaped structure, and since each node represents a smart contract with distinct features, the data are well-suited to apply graph representation learning to obtain graph-level embedding.
The method we employ, graph2vec \cite{narayanan2017graph2vec}, is based on the Weisfeiler-Lehman (WL) method \cite{leman1968reduction} and was subsequently applied in performing graph isomorphism testing \cite{shervashidze2011weisfeiler}.
The graph2vec algorithms leveraging the WL subtree kernel measure graph similarity quantitatively based on the commonality of their subgraph components, asserting that graphs share a higher number of common subgraph components and exhibit higher similarity. 
We apply graph2vec to generate a graph-level embedding space, yielding a singular vector for each building block.
The embedding vectors generated abstract the building block similarity relationships into a generalized representation. This offers interpretability and predictive utility when embedding vectors are used as input for downstream models such as classifiers or clustering algorithms.

\subsection{Node Features}\label{sec:nodefeatures}
In graph2vec, the differentiation of the building block node features directly influences the characterization of the subgraphs, consequently affecting how closely graphs are positioned in the embedding space; graphs containing similar node features in addition to similar subgraphs are placed closer to each other.
With graph2vec, we can provide node features or leave the field empty during the embedding vector generation process.
We utilize various node features to investigate how they affect subgraph similarity in the embedding space. 

\textit{None}:
\label{none_feature}
For the baseline setting, no features are assigned to nodes. In this case, graph2vec will assign the node degree by default. The examination of the embeddings is based only on the building block graph structure without the influence of node-specific information.

\textit{3-class}: 
We assigned features to the building block nodes following the distinction across simplified contract types described in the literature \cite{kitzler2022disentangling} and reported in Section~\ref{sec:bg}: factory deployed contracts (i.e., contracts that generate other contracts), assets, and other contracts.

\textit{Signatures Selectors}: Each node $N_i$ in a building block represents a contract address and therefore contains a list of functions. A function selector can be produced for each function, which refers to the first 4 bytes (8 characters) of the Keccak-256 (SHA3) hash of a function's signature, i.e., the name of the function and its input argument types~\cite{contract_abi}.
We utilized the signature extraction tools\footnote{https://github.com/gsalzer/ethutils/tree/main/doc/fourbytes} suggested by Di Angelo et al. \cite{di2020tokens} for obtaining the selectors, representing the signatures, for every node within each building block across the entire building block dataset.
For each node, we generated a list of function selectors and assigned a unique marker to identical lists, utilizing this marker as the feature for the node.

\textit{Signatures Group}:
\label{function_selector_group}
A potential limit of the \textit{signatures} feature is that two contracts with mostly identical functions but minor variations would be assigned a different marker indicating distinct node features in the above function selectors feature.
To precisely represent a smart contract node features by its functionality regardless of minor functional differences, we categorize contracts into distinct groups by evaluating the pairwise similarities of function selectors using the Jaccard metric.
Each building block will be assigned a marker indicating the function selector (representing the signature) group as its node feature.

\subsection{Building Block Labels}
\label{bb_labels}
To assess the performance of the embedding vectors in tasks that utilize these embeddings as input, we employed the following sources as the building block target label.

\textit{Protocol}: The root node's protocol serves as the target label for the building block.

\textit{Financial Functionality Category (FFC)}: We assign each building block to one of eight categories representing its financial functionality based on its root method name using regular expression patterns with the keywords detailed below in Table \ref{tab:function_keywords}. Regular expressions are used to search for specific keywords within the root method names of the building blocks. For instance, if a root method name contains substrings such as ``deposit'' or ``lend'', it is categorized under the ``Lock Capital'' action. The presence of terms negated by ``NOT'' qualifiers, such as in ``stake NOT unstake'', makes a method only categorized as ``Lock Capital'' if it involves ``stake'' without simultaneous involving ``unstake''.

We note that the information used for the building block target label \textit{Financial Functionality Category} differs from that used as node feature for the \textit{Signatures Selectors} and the \textit{Signatures Group}; indeed, the former uses information from the name of the function invoked only, while the latter two use data of \textit{all} functions of a contract, included their arguments. Moreover, for the target label \textit{Protocol}, we have labels for all 10,000 building blocks. For the \textit{Financial Functionality Category} label, instead, not all building blocks contained one of the regular expressions defined in Table~\ref{tab:function_keywords}; these were categorized as `other' and excluded from the evaluation. 
Also, we have filtered out the building blocks with only one node due to the limited capacity to provide meaningful structural information for clustering and analysis purposes.
Further details on the target labels are reported in Tables \ref{tab:data_labels_distribution_financial_category} and \ref{tab:data_labels_distribution_protocol} in Appendix.

\begin{table}[htbp!]
\centering
\vspace{0.1cm}
\caption{Financial Functionality Category and the associated signature keywords.}
\label{tab:function_keywords}
{\small 
\begin{tabular}{@{}p{3.7cm}p{9cm}p{1cm}@{}}
\toprule
\textbf{Financial functionality category} & \textbf{Keywords} & \textbf{Count} \\ \midrule
Swap & `swap', `exchange' & 4950 \\ \arrayrulecolor{black!30}\midrule
Lock Capital & \begin{tabular}[c]{@{}l@{}}`deposit', `add AND liquidity', `staking', `stake NOT \\ unstake', `lock NOT unlock NOT block', `lend', `collateralize'\end{tabular} & 550 \\ \arrayrulecolor{black!30}\midrule
Redeem or Withdraw & \begin{tabular}[c]{@{}l@{}}`withdraw', `remove AND liquidity',\\  `unstake', `unstaking', `unlock'\end{tabular} & 512 \\ \arrayrulecolor{black!30}\midrule
Borrow & `borrow' & 139 \\ \arrayrulecolor{black!30}\midrule
Get Interest or Rewards & \begin{tabular}[c]{@{}l@{}}`(get OR claim) AND (reward OR fee)', `harvest', `earn'\end{tabular} & 129 \\ \arrayrulecolor{black!30}\midrule
Repay & `repay' & 36 \\ \arrayrulecolor{black!30}\midrule
Governance & `vote' & 16 \\ \arrayrulecolor{black!30}\midrule
Liquidate & `liquidate', `liquidation' & 2 \\ \arrayrulecolor{black!30}\midrule
Others & - & 3666 \\ \bottomrule
& & \\
\end{tabular}
}
\end{table}

\section{Clustering Analysis}
\label{sec:clustering}

In this section, we describe the analysis devised to assess how effectively our embeddings categorize DeFi building blocks into clusters that reflect their financial functionalities, or other information such as the protocol they are associated with.

The procedure is presented in the workflow formulated in \ref{expts_clustering}.
We conducted the analyses using each node feature $\mathcal{F}_{B_i}$ (see Section~\ref{sec:nodefeatures}) and the target building block labels $\mathcal{L}_{B_i}$ (see Section~\ref{bb_labels}), as described in line 3.
We applied the graph2vec algorithm (line 4-8) with an embedding dimension of $\mu = 128$, a learning rate $\gamma = 0.05$, and $e = 100$ epochs to all the 10,000 building blocks. We used a threshold of 1.5 for Jaccard Ward distance when producing \textit{Signatures Group} node features. 
After we obtained an embedding vector for each building block, we applied agglomerative clustering 
\cite{murtagh2014ward}  on all the embeddings.
To determine the optimal clustering distance threshold $\delta$ (line 9), we examined values within the range of [0.6, 1]. The value in this range that yielded the highest V-measure was chosen to have an optimal balance between homogeneity (to which extent each cluster contains only building blocks of a single target label) and completeness (to which extent all building blocks of a given target label are assigned to the same cluster)\footnote{We note that V-Measure is equivalent to Normalized Mutual Information (NMI).}.

\begin{algorithm}[h]
	\DontPrintSemicolon
	\renewcommand{\thealgocf}{Algorithm 1}
	\KwData{A set of building block graphs $\mathcal{B} = \{B_1, B_2, ..., B_n\}$. Parameters: dimension $\mu$, number of epochs $e$, learning rate $\gamma$. Clustering distance threshold: $\delta$}.
	\Begin{
		\ForEach{building block $B_i$ in $\mathcal{B}$}{
			Assign each node in $B_i$ with the feature $\mathcal{F}_{B_i}$, and a set of labels $\mathcal{L}_{B_i}$\;
		}
		
		Initialize matrix $\Phi \in \mathbb{R}^{|\mathcal{B}| \times \mu}$ for building block embeddings\;
		\For{epoch $e = 1$ \KwTo $E$}{
			Shuffle $\mathcal{B}$\;
			\ForEach{building block $B_i$ in $\mathcal{B}$}{
				Update $\Phi(B_i)$ using graph learning model ($B_i$, $\mathcal{F}_{B_i}$, $\mu$, $\gamma$)\;
			}
		}
		
		$\mathcal{C}, \mathcal{L}_{\text{cluster}} \gets \text{agglomerative\_clustering}(\Phi, \delta)$\;
		Assign predicted cluster $\mathcal{C}$ to each $B_i$\;
		
		Evaluation: Calculate Homogeneity (H), Completeness (C), V-measure (V) and Purity (P)  
		using $\mathcal{C}$ and labels $\mathcal{L}_{B_i}$.
	}
	\caption{Building Block Clustering.}
	\label{expts_clustering}
\end{algorithm}

\subsection{Results}

\begin{figure}[h!]
    \centering
    \includegraphics[width=0.99\linewidth]{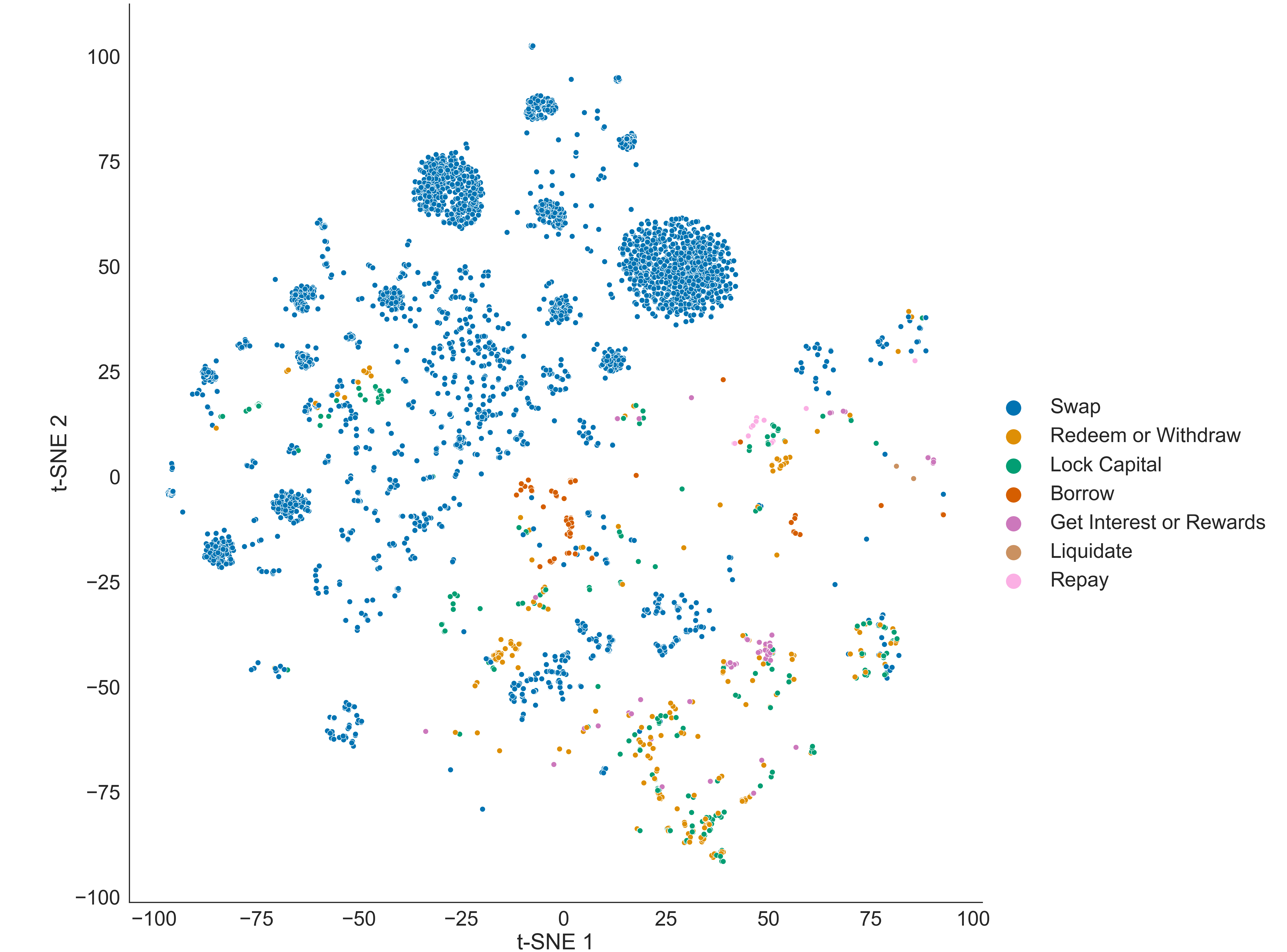}
    \caption{Visualization of the embedding space of building blocks using \textit{Signatures Group} as node feature and \textit{Financial Functionality Category} as label. The building blocks within the Swap financial functionality category are well separated from the other categories and form multiple clusters. Building blocks of certain functionalities, such as `Redeem or Withdraw' and `Lock Capital' stay close, indicating overlapping characteristics despite financial functionality category.}
    \label{fig:financial_categories}
\end{figure}

\begin{figure}[h!]
    \centering
    \includegraphics[width=0.95\linewidth]{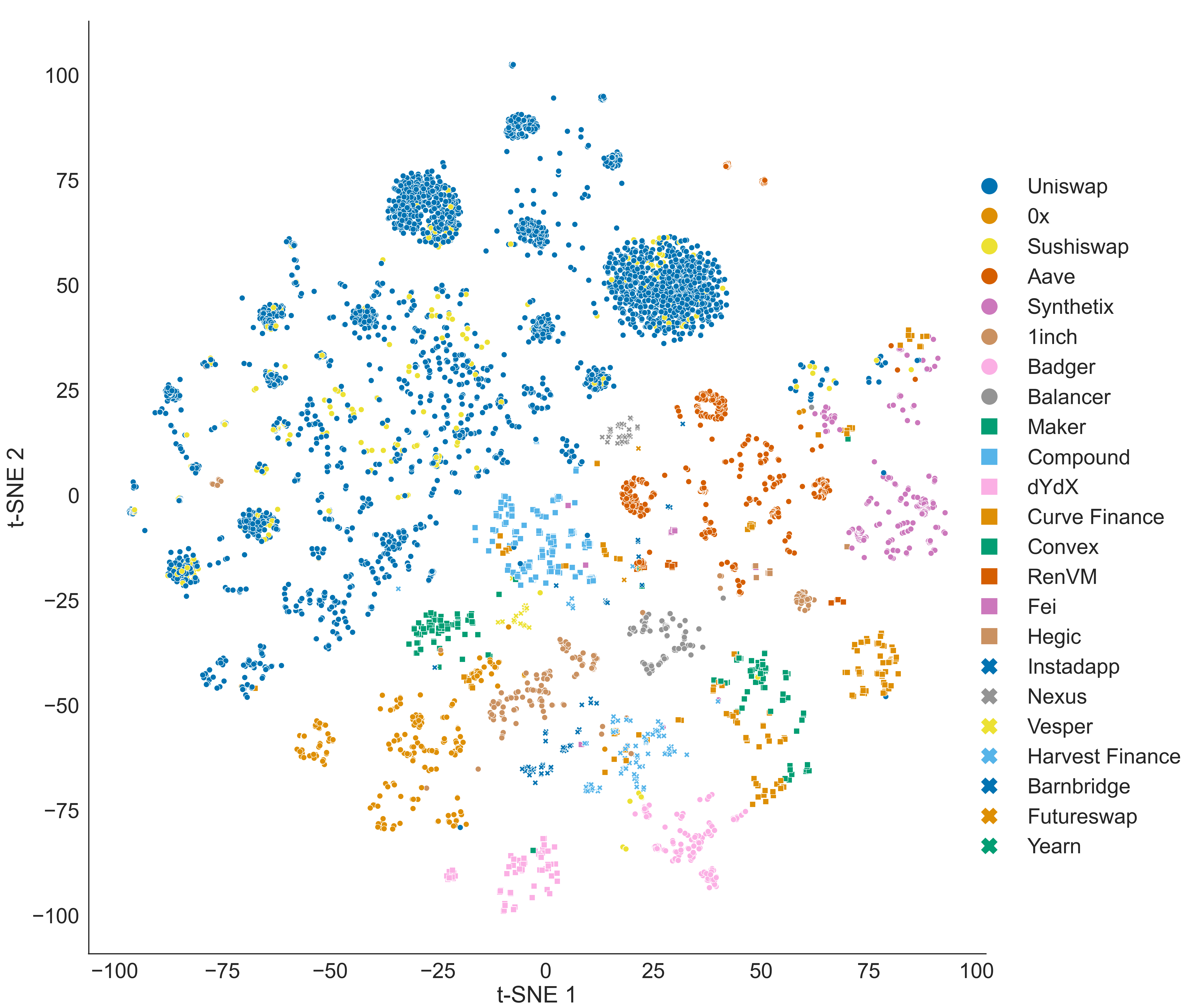}
    \caption{Visualization of the embedding space of building blocks using \textit{Signatures Group} as node feature and \textit{Protocol} as label. Clusters of most protocols are markedly separated, with exceptions such as \emph{Uniswap} and \emph{Sushiswap}; this reflects well that the latter is a forked project of the former.}
    \label{fig:protocol_categories}
\end{figure}

We evaluate the clustering performance by computing the homogeneity, completeness, V-measure, and purity over both two target labels and the four node features defined in Section \ref{buildling_block_embedding}.
Table \ref{tab:clustering_results} reports the results of our analyses.
We highlight in gray the best-performing specifications. 
Interestingly, we find that in all cases (apart from one) the node feature \textit{Signatures Group} yields the best results. 
This is in line with our expectations since this feature is the most advanced and captures best the characteristics of the building blocks; the more information incorporated into the features, the better the results become.
We first focus on the \textit{Financial Functionality Category} target label.
In this specification, the outcomes yielded the highest value among the feature sets for purity with $.888$ (using \textit{Signatures Group} as node feature). To further interpret these results, we use t-SNE and reduce the embedding space into two dimensions. Figure \ref{fig:financial_categories} shows the embedding space of building blocks using \textit{Signatures Group} as the node feature and \textit{Protocol} as the label.
In the figure, each dot represents a building block's embedding vector, with dimension reduced to 2D using t-SNE and its colour indicating the corresponding building block label.
We find that building blocks associated with swapping functions, which are the vast majority of building blocks, are clustered close to each other. 
On the other side,  completeness and V-measure are relatively low. This could be caused by clusters, such as `Redeem or Withdraw' and `Lock Capital', being not well separated. However, this shows an interesting pattern, as the two functionalities are actually reciprocal to each other. 
In summary, the results suggest that the clustering categorizes DeFi building blocks into clusters where a substantial portion aligns with their distinct financial functionalities. 
Next, we look at the \textit{Protocol} target labels and investigate how the performance changes.
As Table~\ref{tab:clustering_results} shows, the best-performance purity of $.864$ is comparable to the one of FFC, whilst all other measures are relatively higher.
Figure \ref{fig:protocol_categories} illustrates the observed separation by visualizing the embedding space using t-SNE dimension reduction (and additionally using different marker shapes to differentiate similar colours).
The formed clusters show clear overall separation based on protocol: building blocks within the same protocol tend to share more common interaction patterns, characterized by similar sub-graph structures and features, whilst they are distinct from building blocks of other protocols.
Notably, an interesting observation is that building blocks from \emph{Uniswap} and \emph{Sushiswap} exhibit close proximity, indicating overlapping functionalities as expected, since \emph{Sushiswap} is a fork of \emph{Uniswap} \cite{sushiswap_whitepaper}.
In conclusion, we find that the graph embedding method works better in separating building blocks associated with the same protocol in comparison with FFC.

\begin{table*}[h!]
\centering

\setlength{\extrarowheight}{0pt}
\addtolength{\extrarowheight}{\aboverulesep}
\addtolength{\extrarowheight}{\belowrulesep}
\setlength{\aboverulesep}{0pt}
\setlength{\belowrulesep}{0pt}

\caption{Clustering results on building blocks with combinations of node features and building block labels. The best results for each target label are highlighted through gray shading, indicating that the \textit{Signature Group} node feature produced the optimal clustering outcome evaluated by both target labels.}
\resizebox{1.0\columnwidth}{!}{%
\begin{tabular}{llcccccc} 
\toprule
Target Label & Node Feature & $\delta$ & Cluster & Homogeneity & Completeness & V-measure & Purity \\ 
\cline{1-2}\cmidrule{3-8}
\multirow{4}{*}{Protocol} & None & 0.64 & 294 & 0.371 & 0.202 & 0.262 & 0.582 \\
 & 3-class & 0.61 & 371 & 0.596 & 0.284 & 0.385 & 0.700 \\
 & Signatures Selectors & 0.70 & 223 & 0.822 & {\cellcolor[rgb]{0.8,0.8,0.8}}0.436 & 0.570 & 0.860 \\
 & Signatures Group & 0.64 & 251 & {\cellcolor[rgb]{0.8,0.8,0.8}}0.838 & 0.432 & {\cellcolor[rgb]{0.8,0.8,0.8}}0.571 & {\cellcolor[rgb]{0.8,0.8,0.8}}0.864 \\ 
\hline
\multirow{4}{*}{\begin{tabular}[c]{@{}l@{}}Financial Functionality\\Category\end{tabular}} & None & 0.66 & 279 & 0.463 & 0.090 & 0.160 & 0.849 \\
 & 3-class & 0.70 & 280 & 0.594 & 0.120 & 0.200 & 0.856 \\
 & Signatures Selectors & 0.60 & 283 & 0.689 & 0.135 & 0.225 & 0.887 \\
 & Signatures Group & 0.62 & 265 & {\cellcolor[rgb]{0.8,0.8,0.8}}0.706 & {\cellcolor[rgb]{0.8,0.8,0.8}}0.144 & {\cellcolor[rgb]{0.8,0.8,0.8}}0.239 & {\cellcolor[rgb]{0.8,0.8,0.8}}0.888 \\
\bottomrule
\end{tabular}
}
\label{tab:clustering_results}
\end{table*}

\clearpage

\subsection{Further Insights}
\begin{figure}[t]
    \centering
    \vspace{0.35cm}
    \includegraphics[width=1\textwidth, scale=1]{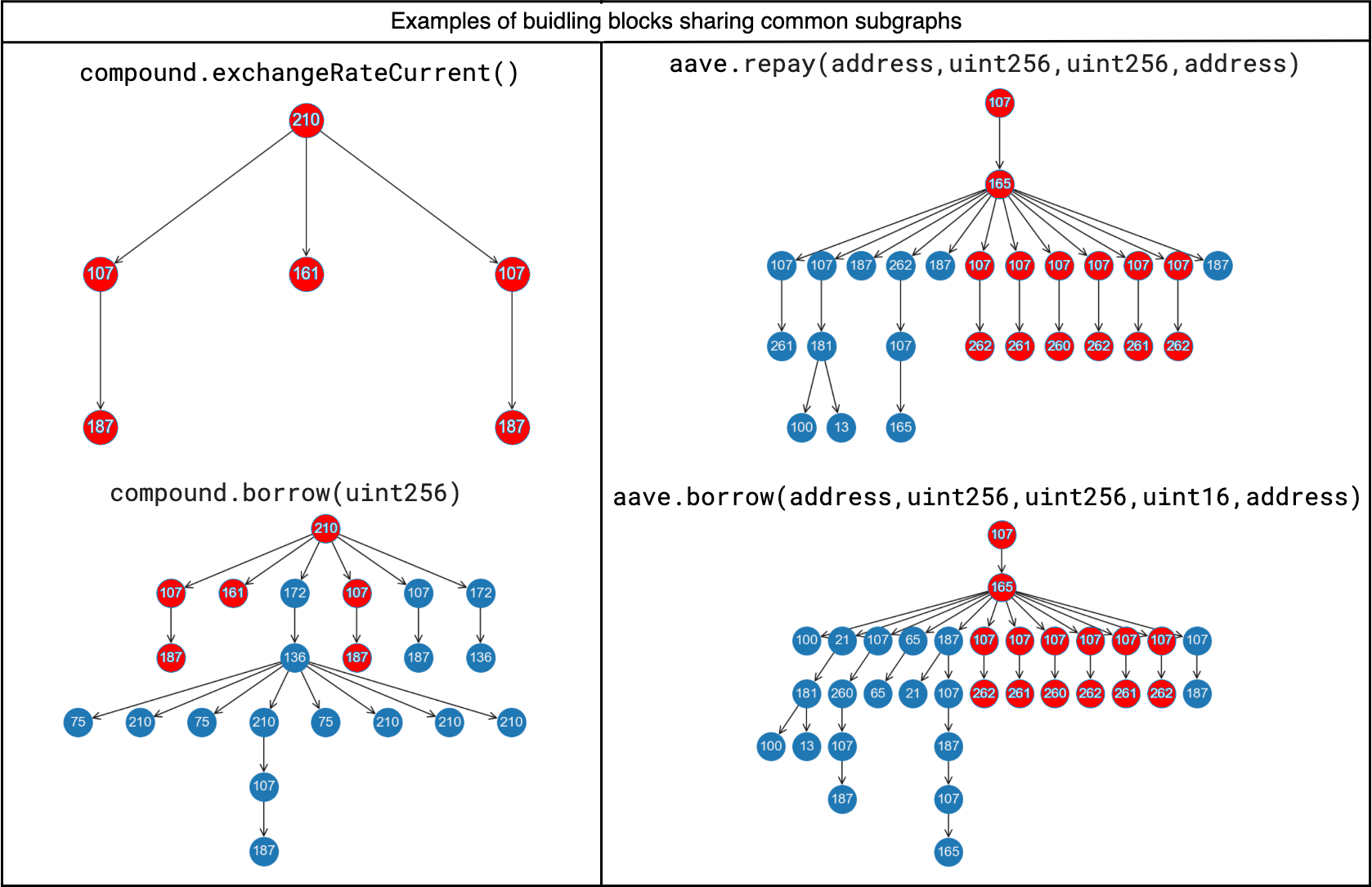} 
    \caption{Examples of building blocks with shared common subgraphs highlighted in red. Numeric values in nodes represent distinct \textit{Signature Group} node features. Building blocks of same protocols, while differing in financial functionalities, can either contain the entire subgraph of another (left) or share common subgraphs (right), suggesting the presence of protocol-specific patterns reuse. }    
    \label{fig:non_majority_llustration}
    \vspace{0.3cm}
\end{figure}

When using FFC as target label, we observed a more fragmented scenario in clusters (see Figure \ref{fig:financial_categories}) in contrast to \emph{Protocol} as target. 
To look for plausible explanations and provide a deeper understanding of the method's performance, we therefore investigate more closely such differences.
In Figure~\ref{fig:non_majority_llustration}, we report illustrative examples of building blocks associated with the same protocol, that either contain the entire subgraph of another (left) or share common subgraphs (right).
Numbers inside nodes represent the IDs of the function signature group features.
Common transaction patterns are highlighted in red.
On the left, the 
building block `borrow' of the lending protocol \emph{Compound} contains the entire graph of `exchangeRateCurrent', although the financial functionalities differ. This re-used subgraph suggests that there likely are internal mechanisms that are not specific to the financial functionality, but instead are protocol-specific patterns. 
Also, on the right illustration we find for the lending protocol \emph{Aave} an overlapping pattern, here with a common subgraph within `repay` and 'borrow'. Both examples are indicators for the existence of protocol-specific patterns, that are re-used across them, without necessarily being part of the financial functionality. 
These could be complementary CA-calls to offer the financial services, e.g. get token exchange rates.
Consequentially, these common subgraphs might reason higher measures of the protocol level clustering, compared to the financial functionalities. 

\section{Discussion and Conclusions}
\label{sec:conclusions}

In this paper, we applied graph representation learning and agglomerative algorithms to cluster basic financial services offered by Decentralized Finance (DeFi) protocols, also called DeFi \textit{building blocks}. We measured their similarity based on the subgraph structure of their corresponding DeFi transactions and by exploiting the smart contract attributes as node features. 
Using as target label information on the protocols and the financial categories the building blocks are associated with, we can assess the effectiveness of our method and find that we are able to cluster building blocks associated with the same financial category with purity and V-measure respectively as high as $.888$ and $.571$.

With our method, we are able to provide a broad overview of the DeFi building blocks that offer similar financial functions. As an example, \emph{Uniswap} and \emph{Sushiswap} building blocks are in their very near proximity and often overlap, showing that our approach can identify that the latter is a forked project of the former; for other protocols, it is instead visible a clear distinction of protocol-specific clusters, whereby further categorizations of similar clusters take place within protocols.
Our study also provides a broad overview and categorization of a large variety of the most utilized building blocks in DeFi. This mapping might foster interoperability and facilitate the implementation of novel smart contracts that can interact with many existing building blocks.

The devised methodology could benefit from implementing the following improvements. 
First, we only focus on transaction data and do not take additional blockchain-related event data into account to extract, e.g. transfer patterns.
Also, the dataset we used is limited in the time period analyzed and protocols included. Whilst the main purpose of the paper is to devise a novel approach to cluster similar building blocks, the extension to a longer time frame could provide further insights and also enable a temporal analysis. 
Finally, we currently focus only on a limited number of machine learning and graph representation algorithms. As a next step, we propose to complement them with additional analyses to cluster and categorize the DeFi functionalities. 

To improve the clustering, future work should focus on handling protocol-specific transaction patterns. 
Also, a very promising approach to improve the methodology could entail the use of Large Language Models (LLMs), for instance, to extract smart contract information as a node feature. Such applications can be extended further, for example, to explore possible financial functionalities of building blocks emerging from new transactions, and predict their functionality based on their structure and attributes.

In conclusion, we believe that our work can help provide a tool for identifying common patterns across DeFi functionalities and sheds light on the landscape of services offered by DeFi protocols. 

\newpage

\section*{Author Contributions}
JL developed the algorithm and conducted the analyses; SK and PS extracted the data and conceptualized the research questions; all authors interpreted the results and contributed equally to the writing of the paper.

\bibliographystyle{naturemag}
\bibliography{refs.bib}

\newpage

\begin{appendix}
\section{Appendix}
\setcounter{section}{0}
\renewcommand{\thesection}{\Alph{section}}

\subsubsection*{Details on Building Block Labels}

Tables \ref{tab:data_labels_distribution_financial_category} and \ref{tab:data_labels_distribution_protocol} provide additional details on the target labels used to evaluate the clustering embeddings.

\vspace{10px}

\begin{table}[htbp!]
\centering
\caption{Label Distribution within \textit{Financial Functionality Category}. AF: After filtering.}
{\small
\begin{tabular}{lll}
\hline
\textbf{Financial Functionality Category}  & \textbf{Count} & \textbf{AF} \\ \hline
Swap & 4950 & 4930 \\
Lock Capital & 550 & 543 \\
Redeem/Withdraw & 512 & 506 \\
Interest/Rewards & 129 & 128 \\
Borrow & 139 & 139 \\
Repay & 36 & 36 \\
Governance & 16 & 8 \\
Liquidate & 2 & 2 \\
Others & 3666 & 3485 \\ 
Total & 10000 & 9877 \\ \hline
\end{tabular}
} 
\label{tab:data_labels_distribution_financial_category}
\vspace{0.8cm}
\end{table}

\begin{table}[ht]
\centering
\caption{Label Distribution within \textit{Protocol}. AF: After filtering.}
{\small 
\begin{tabular}{lllllllll} 
\toprule
\textbf{Protocol} & \textbf{Count} & \textbf{\textbf{AF}} & \textbf{Protocol} & \textbf{Count} & \textbf{AF} & \textbf{Protocol} & \textbf{Count} & \textbf{AF} \\ \hline
Uniswap & 4475 & 4448 & Badger & 283 & 281 & RenVM & 61 & 58 \\
Aave & 795 & 795 & dYdX & 244 & 244 & Vesper & 44 & 42 \\
0x & 618 & 606 & Harvest Finance & 238 & 234 & Hegic & 36 & 34 \\
Sushiswap & 532 & 517 & Balancer & 207 & 202 & Fei & 30 & 25 \\
Synthetix & 500 & 497 & Convex & 197 & 195 & Barnbridge & 26 & 24 \\
Curve Finance & 495 & 489 & Maker & 194 & 182 & Futureswap & 11 & 11 \\
Compound & 424 & 420 & Instadapp & 148 & 148 & Yearn & 5 & 2 \\
1inch & 346 & 334 & Nexus & 91 & 89 & Total & 10000 & 9877 \\
\bottomrule
\end{tabular}
}
\vspace{0.8cm}
\label{tab:data_labels_distribution_protocol}
\end{table}

\end{appendix}

\end{document}